\documentclass[11pt]{article}

\usepackage{tikz}
\usetikzlibrary{shapes}
\usepackage{algorithm}
\usepackage{program}
\usepackage{algpseudocode}
\usepackage{pifont}

\usepackage{epsfig}
\usepackage{tabularx}
\usepackage{multirow}
\usepackage{graphicx}

\newtheorem{theorem}{Theorem}

\title{Computing Pretropisms for the Cyclic $n$-Roots 
Problem\thanks{This material is based upon work 
supported by the National Science Foundation under Grant No. 1440534.}}

\author{Jeff Sommars\thanks{Department of Mathematics, Statistics, and Computer Science,
        University of Illinois at Chicago, {\tt sommars1@uic.edu}}
        \and
        Jan Verschelde\thanks{Department of Mathematics, Statistics, and Computer Science,
        University of Illinois at Chicago, {\tt janv@uic.edu}.}}

\begin{document}
\maketitle

\begin{abstract}
The cyclic $n$-roots problem is an important benchmark problem
for polynomial system solvers.  We consider the pruning of cone
intersections for a polyhedral method to compute series
for the solution curves.
\end{abstract}

\section{Introduction}

The cyclic $n$-roots problem asks for the solutions
of a polynomial system, commonly formulated as
\begin{equation} \label{eqcyclicsys}
   \begin{cases}
   x_{0}+x_{1}+ \cdots +x_{n-1}=0 \\
   i = 2, 4, \ldots, n-1: 
    \displaystyle\sum_{j=0}^{n-1} ~ \prod_{k=j}^{j+i-1}
    x_{k~{\rm mod}~n}=0 \\
   x_{0}x_{1}x_{2} \cdots x_{n-1} - 1 = 0. \\
\end{cases}
\end{equation}
This problem is important in the study of biunimodular vectors,
a notion that traces back to Gauss, as stated in~\cite{FR15}.
In~\cite{Bac89}, Backelin showed that if $n$ has a divisor that
is a square, i.e. if $d^2$ divides $n$ for $d \geq 2$, then
there are infinitely many cyclic $n$-roots.
The conjecture of Bj{\"{o}}rck 
and Saffari~\cite{BS95}, \cite[Conjecture~1.1]{FR15}
is that if $n$ is not divisible by a square, 
then the set of cyclic $n$-roots is finite.

As shown in~\cite{AV12},
the result of Backelin can be recovered by polyhedral methods.
Polyhedral methods to solve a polynomial system consider the
Newton polytopes of the polynomials in the system.
The {\em Newton polytope} of a polynomial in several variables
is the convex hull of the exponent tuples of the monomials that
appear with nonzero coefficient in the polynomial.
Looking for positive dimensional solution sets,
we look for series developments of the solutions,
and in particular we look for Puiseux series.
The leading exponents of Puiseux series are called {\em tropisms}.
A {\em pretropism} is a vector that forms
the minimal inner product with a face of every one of the given
polytopes, where none of the faces are 0-faces.
Pretropisms are candidates for being tropisms, but 
not every pretropism is a tropism, as pretropisms depend only on
the Newton polytopes of the system, see e.g.~\cite{MS15}
for a mathematical background on tropical algebraic geometry.

Our problem can thus be stated as follows.
Given a tuple of Newton polytopes, compute all pretropisms.
In~\cite{SV15} we examined the case where all polytopes
are in general position with respect to each other.
In this paper
we focus on the Newton polytopes of the cyclic $n$-roots problem.

\noindent {\bf Prior and related work.}
In~\cite{BJSST07}, the computation of pretropisms is defined as the 
common refinement of the normal fans of the Newton polytopes~\cite{Zie95}.
The software Gfan~\cite{Jen08} relies on cddlib~\cite{FP96}
in its application of reverse search algorithms~\cite{AF92}.

\section{Pruning Cone Intersections}

\begin{figure}[h]
\begin{center}
\begin{picture}(200,240)(0,0)
\put(0,125){
\begin{tikzpicture}
\node (root) at (3,6) [rectangle] {};
\node (A) at (1.5,4.5)  [rectangle]{A};
\node (B) at (3, 4.5) [rectangle]{B};
\node (C) at (4.5,4.5)  [rectangle]{C};
\draw (root) edge (A)  (root) edge (B) (root) edge (C);

\node (D) at (0,3.5) {D};
\node (E) at (0.5,3.5) {E};
\node (F) at (1,3.5) {F};
\node (G) at (1.5,3.5) {G};
\draw (A) edge (D)  (A) edge (E)  (A) edge (F)  (A) edge (G);

\node (H) at (2.5,3.5) {H};
\node (I) at (3,3.5) {I};
\node (FF) at (3.5,3.5) {F};
\draw (B) edge (H)  (B) edge (I)  (B) edge (FF);

\node (J) at (4.5,3.5) {J};
\node (HH) at (5,3.5) {H};
\node (GG) at (5.5,3.5) {G};
\draw (C) edge (J)  (C) edge (HH) (C) edge (GG);

\node (K) at  (-0.5,2.5) {K};
\node (L) at  (0,2.5) {L};
\draw (D) edge (K) (D) edge (L);

\node (M) at  (0.4,2.5) {M};
\node (N) at  (0.7,2.5) {N};
\node (O) at  (1.0,2.5) {O};
\node (P) at  (1.3,2.5) {P};
\draw (F) edge (M) (F) edge (N) (F) edge (O) (F) edge (P);

\node (Q) at (1.7,2.5) {Q};
\draw (G) edge (Q);

\node (R) at  (2.2,2.5) {R};
\node (S) at  (2.5,2.5) {S};
\draw (H) edge (R) (H) edge (S);

\node (T) at  (2.8,2.5) {T};
\node (U) at  (3.1,2.5) {U};
\draw (I) edge (T) (I) edge (U);

\node (MM) at  (3.4,2.5) {M};
\node (NN) at  (3.7,2.5) {N};
\node (OO) at  (4.0,2.5) {O};
\node (PP) at  (4.3,2.5) {P};
\draw (FF) edge (MM) (FF) edge (NN) (FF) edge (OO) (FF) edge (PP);

\node (RR) at  (4.9,2.5) {R};
\node (SS) at  (5.3,2.5) {S};
\draw (HH) edge (RR) (HH) edge (SS);

\node (QQ) at (5.8,2.5) {Q};
\draw (GG) edge (QQ);
\end{tikzpicture}
}
\put(0,0){
\begin{tikzpicture}
\node (root) at (3,6) [rectangle] {};
\node (A) at (1.5,4.5)  [rectangle]{A};
\node (B) at (3, 4.5) [rectangle]{B};
\node (C) at (4.5,4.5)  [rectangle]{C};
\draw (root) edge (A)  (root) edge (B) (root) edge (C);

\node (D) at (0,3.5) {D};
\node (E) at (0.5,3.5) {E};
\node (F) at (1,3.5) {F};
\node (G) at (1.5,3.5) {G};
\draw (A) edge (D)  (A) edge (E)  (A) edge (F)  (A) edge (G);

\node (H) at (2.5,3.5) {H};
\node (I) at (3,3.5) {I};
\node (FF) at (3.5,3.5) {F};
\draw (B) edge (H)  (B) edge (I)  (B) edge (FF);

\node (J) at (4.5,3.5) {J};
\node (HH) at (5,3.5) {H};
\node (GG) at (5.5,3.5) {G};
\draw (C) edge (J)  (C) edge (HH) (C) edge (GG);

\node (K) at  (-0.5,2.5) {K};
\node (L) at  (0,2.5) {L};
\draw (D) edge (K) (D) edge (L);

\node (M) at  (0.4,2.5) {M};
\node (N) at  (0.7,2.5) {N};
\node (O) at  (1.0,2.5) {O};
\node (P) at  (1.3,2.5) {P};
\draw (F) edge (M) (F) edge (N) (F) edge (O) (F) edge (P);

\node (Q) at (1.7,2.5) {Q};
\draw (G) edge (Q);

\node (R) at  (2.2,2.5) {R};
\node (S) at  (2.5,2.5) {S};
\draw (H) edge (R) (H) edge (S);

\node (T) at  (2.8,2.5) {T};
\node (U) at  (3.1,2.5) {U};
\draw (I) edge (T) (I) edge (U);

\end{tikzpicture}
}
\end{picture}
\caption{Nodes A, B, C represent cones to $P_1$.
Intersections of those cones with the cones of $P_2$
are represented by nodes D through J.
Duplicate nodes are removed from the second tree. }
\label{figconegraphs}
\end{center}
\end{figure}
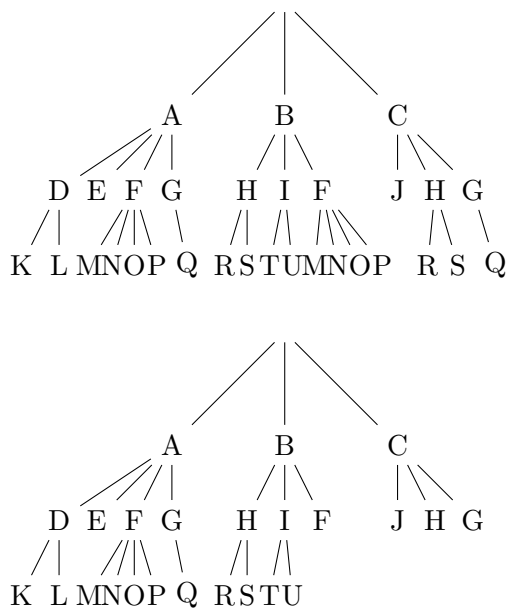

To introduce our algorithms, consider Figure~\ref{figconegraphs}.
For three Newton polytopes $(P_1, P_2, P_3)$,
the leaves of the trees represent cones of pretropisms.
Nodes without children that are not leaves correspond to
cone intersections that contain only the zero dimensional
cone.

The removal of duplicate nodes eliminates many cone intersections
at deeper levels in the tree. 

\section{Algorithms}

Our algorithm takes as input the edge skeletons of a set of Newton
polytopes; the edge skeleton of a polytope can be computed by a 
polynomial-time algorithm, presented in~\cite{EFG16}. In our 
implementation, we use edge objects that have vertices, 
references to their neighboring edges, and the cone of the set of 
inner normals of all of the facets on which the edge rests.
Note that since the cyclic-$n$ polytopes are not all full dimensional, 
we included generating rays of the lineality spaces as needed.

Algorithm~\ref{horizontalprune} sketches the outline of the algorithm
to compute all pretropisms of a tuple of $n$ polytopes.
Along the lines of the gift wrapping algorithm,
for every edge of the first polytope we take the plane that contains
this edge and consider where this plane touches the second polytope.
Algorithm~\ref{algedgeskeleton} starts exploring the edge skeleton defined
by the edges connected to the vertices in this touching plane.

\begin{algorithm}[hbt]
\begin{algorithmic}[1]
\caption{Explores the skeleton of edges to find pretropisms
         of a polytope $P$ and a cone $C$.}
\label{algedgeskeleton}
\Function{ExploreEdgeSkeleton}{$P$, $C$}
\State $r$ := a random ray inside $C$
\State $m$ := $\min\{ \langle a, r \rangle, a \in P \}$
\State ${\rm in}_{r}(P)$ := $\{ a \in P, \langle a , r \rangle = m \}$	
\State EdgesToTest := edges $e$ of $P$: $e \cap {\rm in}_{r}(P) \not= \emptyset$
\State Cones := $\emptyset$
\State TestedEdges := $\emptyset$
\While{EdgesToTest $\not= \emptyset$}
\State $E$ := pop an edge from EdgesToTest
\State $C_E$ := normal cone to $E$
\State ShouldAddCone := False
\If{$C_E$ contains $C$}
\State ConeToAdd := $C$
\State ShouldAddCone := True
\ElsIf
{$C \cap C_E \not= \{ 0 \}$}
\State ConeToAdd := $C \cap C_E$
\State ShouldAddCone := True
\EndIf

\If{ShouldAddCone}
\State Cones := Cones $\cup~$ConeToAdd
\State Edges := Edges $\cup~E$
\For{each neighboring edge $e$ of $E$}
\If{$e \not\in$ TestedEdges}
\State EdgesToTest := EdgesToTest$\cup e$
\EndIf
\EndFor
\EndIf
\State TestedEdges := TestedEdges $\cup~E$
\EndWhile
\State \Return Cones
\EndFunction
\end{algorithmic}
\end{algorithm}

The exploration of the neighboring edges corresponds to tilting
the ray~$r$ in Algorithm~\ref{algedgeskeleton}, as in rotating a 
hyperplane in the gift wrapping method.
One may wonder why the exploration of the edge skeleton in 
Algorithm~\ref{algedgeskeleton} needs to continue after 
the statement on line~5. This is because the cone $C$ has
the potential to intersect many cones in $P$, particularly if $P$ has
small cones. Furthermore it is reasonable to wonder why we 
bother checking cone containment when computing the intersection
of two cones provides more useful information. Checking cone
containment means checking if each of the generators of $C$ is contained
in $C_E$, which is a far less computationally expensive operation
than computing the intersection of two cones.

In the Newton-Puiseux algorithm to compute series expansions, we are 
interested only in the edges on the lower hull of the Newton polytope,
i.e. those edges that have an upward pointing inner normal~\cite{Wal50}.
For Puiseux for space curves, the expansions are normalized so that
the first exponent in the tropism is positive.
Algorithm~\ref{horizontalprune} is then adjusted so that
calls to the edge skeleton computation of Algorithm~\ref{algedgeskeleton}
are made with rays that have a first component that is positive.

\begin{algorithm}[hbt]
\caption{Finds pretropisms for a given tuple of polytopes
$(P_1, P_2, \ldots, P_n)$.}
\label{horizontalprune}
\begin{algorithmic}[1]
\Function{FindPretropisms}{$P_1$, $P_2, \ldots,$ $P_{n}$}

\State Cones := set of normal cones to edges in $P_1$
\For{i := 2 to n}
\State NewCones := $\emptyset$
\For{Cone in Cones}
\State NewCones := NewCones $\cup$ 
ExploreEdgeSkeleton($P_i$, Cone)
\EndFor
\State Cones := NewCones
\EndFor
\State Pretropisms := set of generating rays for each cone in Cones
\State \Return Pretropisms
\EndFunction
\end{algorithmic}
\end{algorithm}

\subsection{Correctness}
To see that this algorithm will do what it claims,
we must define an additional term. A {\em pretropism graph} is
the set of edges for a polytope that have normal cones
intersecting a given cone. We will now justify why 
the cones output by Algorithm~\ref{algedgeskeleton} correspond
to the set of cones that live on a pretropism graph.

\begin{theorem} \label{theorem1}
Pretropism graphs are connected graphs.
\end{theorem}
\noindent {\em Proof.}
Let $C$ be a cone, and let $P$ be a polytope with edges $e_1, e_2$ such that
they are in the pretropism graph of $C$.
Let $C_1$ be the cone of the intersection of the normal cone of
$e_1$ with $C$, and let $C_2$ be the cone of the intersection 
of the normal cone of $e_2$ with $C$.
If we can show that there exists a path between $e_1$ and $e_2$
that remains in the pretropism graph, then the result will follow.

Let $n_1$ be a normal to $e_1$ that is also in $C_1$ 
and let $n_2$ be a normal to $e_2$ that is also in $C_2$.
Set $n = tn_1 + (1-t)n_2$ where $0 \le t \le 1$. 
Consider varying $t$ from 0 to 1; 
this creates the cone $C_n$, a cone which must 
lie within $C$, as both $n_1$ and $n_2$ lie in that cone. 
As $n$ moves from 0 to 1, it will progressively intersect new faces of $P$
that have all of their edges in the pretropism graph. Eventually, this 
process terminates when we reach $e_2$, and we have constructed
a path from $e_1$ to $e_2$. Since a path
always exists, we can conclude that pretropism graphs are 
connected graphs.
~\qed

Since pretropism graphs are connected, Algorithm~\ref{algedgeskeleton}
will find all cones of edges on the pretropism graph. In 
Algorithm~\ref{horizontalprune}, we repeatedly explore the edge skeleton
of polytope $P_i$, and use the pruned set of cones to explore $P_{i+1}$. From
this, it is clear that Algorithm~\ref{horizontalprune} will suffice
to find all pretropisms.

\section{Comparison With Our Previous Algorithm}

Our algorithm in~\cite{SV15} restricted the pruning of the cone 
intersections in a vertical fashion: nodes in the tree with
cone intersections that yield only $\{ 0 \}$ will not have any children.
That algorithm works well for polytopes with randomly generated coordinates.

In this paper we consider polytopes that are not in generic position.
In this situation, intersecting normal cones to edges may lead to cones
of normals of higher dimensional cones.  At the same level in the tree
we can then have duplicate cones or cones that are contained in other cones.
In those cases were one cone is contained in another, the smaller cone
can be pruned from the tree.  We call this type of pruning horizontal pruning.
For generic polytopes horizontal pruning would not reduce
the number of cone intersections.  However, in special
cases like the cyclic $n$-root problem, there is the potential to dramatically
reduce the scope of the problem through horizontal pruning.

To illustrate horizontal pruning, consider Figure~\ref{figconegraphs}.
These graphs illustrate computing the pretropisms for three fictitious, 
non-generic polytopes $P_1$, $P_2$, $P_3$ with the two distinct algorithms.
Nodes A, B, C represent the cones of the edges of $P_1$,
the row below that represents the resulting cones from performing
Algorithm~\ref{algedgeskeleton} with $P_2$ and A, B, or C.
The row below that varies in the two figures.
In the tree at the top of Figure~\ref{figconegraphs},
the process iterates and Algorithm~\ref{algedgeskeleton} is performed
with $P_3$ and each of the input cones D through J. 
The tree at the bottom of Figure~\ref{figconegraphs}
shows how the horizontal pruning has the potential to improve 
over the previous algorithm. Since there are duplicate nodes for F, G, and H, 
each of these paths only needs to be followed once. Though this does
not lead to dramatic improvements in this fictitious case,
as the number of polytopes increases, the benefit of pruning compounds.

\section{Computational Experiments}

We developed a preliminary version of Algorithm~\ref{horizontalprune}
in Sage~\cite{Sage}, using its modules for lattice polytopes~\cite{Nov11}, 
and polyhedral cones~\cite{BH11}; 
Sage uses PPL~\cite{BHZ08} to compute cone intersections.
Our preliminary code is available
at https://github.com/sommars/GiftWrap. 
We ran the code on a Red Hat Enterprise Linux workstation of Microway,
with Intel Xeon E5-2670 processors at 2.6~GHz.

Instead of directly calculating the pretropisms of the Newton polytopes
of the cyclic $n$-root problem,
we chose to calculate pretropisms of the reduced cyclic $n$-root problem. 
This reformulation~\cite{Emi94} is obtained by performing the substitution 
$x_i = \frac{y_i}{y_0}$ for $i = 0 \dots n-1$. 
Clearing the denominator of each equation leaves the first $n-1$ equations 
as polynomials in $y_1, \dots y_{n-1}$. 
We compute pretropisms of the Newton polytopes
of these $n-1$ equations because they yield meaningful sets of pretropisms.
Calculating with the reduced cyclic $n$-roots problem 
has the benefit of removing
much of the symmetry present in the standard cyclic $n$-roots problem, 
as well
as decreasing the ambient dimension by one. 
Unlike the standard cyclic $n$-roots problem,
some of the polytopes of the reduced cyclic $n$-roots problem
are full dimensional,
which leads to calculation speed ups. A simple transformation can be
performed on the pretropisms we calculate of reduced cyclic $n$-root problem 
to convert them to the pretropisms of cyclic $n$-root problem, so calculating
the pretropisms of reduced cyclic $n$-roots problem is equivalent to 
calculating the pretropisms of the cyclic $n$-roots problem.

In Table~\ref{conetable},
we have recorded the number of cone intersections performed
and the number of times cone containments let us avoid performing additional
intersections for each of the reduced cyclic $n$-root systems with $n \le 10$. 
Table~\ref{conetable} also contains a comparison between the two sums, which
acts as a way of evaluating the quality of the algorithms. We consider
the unit of work of each algorithm to be the total number of intersections
performed, as that is the constraining part of the algorithm.
As $n$ increases, the ratio tends to increase as well,
demonstrating that Algorithm~\ref{horizontalprune}
represents a substantial improvement over our previous algorithm.
We expect that the result would become even more dramatic with higher $n$, but
in our testing, 
our previous algorithm was too inefficient to terminate for $n > 10$.

\begin{table*}[hbt]
\centering
\resizebox{\columnwidth}{!}{%
  \begin{tabular}{|r||r|r|r||r|r|r||r|} \hline
    $n$ & intersections & containments & sum & intersections & containments & sum & ratio\\
\hline
     4 &  63 & 2 & 65 & 54 & 2 &  56 & 1.16071 \\ \hline
     5   &     750 &  20 & 770  & 395   &  5 &  400 & 1.92500 \\ \hline 
     6  &     4,531 & 1,232 & 5,763 & 2,982  &  291 & 3,273 & 1.76076 \\ \hline
     7  &    105,982  & 5,767 & 111,749 & 18,798   &  343 & 19,141 & 5.83820 \\ \hline
     8  &    479,640 & 181,507 & 661,147 & 145,125  & 3,922 & 149,047 & 4.43582 \\ \hline
     9  &   9,232,384 & 1,993,049 & 11,225,433 & 1,101,563  & 16,313 & 1,117,876  & 10.04175 \\ \hline
     10  &  70,026,302 & 23,838,851 & 93,865,153 & 8,846,353  & 165,203 & 9,011,556  & 10.41608 \\ \hline
  \end{tabular}%
}
\caption{Columns two through four contain results when our previous algorithm is
applied to the reduced cyclic $n$-roots problem while columns five through seven
contain the results of Algorithm~\ref{horizontalprune}. The final column
represents the ratio of the previous sum to the sum of Algorithm~\ref{horizontalprune}.}
\label{conetable}
\end{table*}

\section{Comparison with Gfan}

As we reported in~\cite{SV15},
on randomly generated polytopes, our code was competitive 
with Gfan~\cite{Jen08}.
Although the additional pruning criteria presented in this paper
are promising, on the specific cyclic $n$-roots problem,
our Python prototype is not as good as the compiled code of Gfan.
Our code tends to be slower by a factor of two, but we hope
to be more competitive if we improve our ability to exploit
the symmetry of the polytopes.

The computational complexity is such that high level parallelism
is effective. Instead of iterating through all of the cones
from line 5 of Algorithm~\ref{horizontalprune}, we can create a
queue of them and then perform Algorithm~\ref{algedgeskeleton}.
We then initialize some number
of processes and give them successive cones from the queue
until the queue is empty. Once the queue is empty, the resulting
cones are pruned and combined and the algorithm iterates.
We have incorporated this parallelism into our prototype Sage code.



\small 
\bibliographystyle{abbrv}

\end{document}